# Title: Formation of stable aggregates by fluid-assembled solid bridges

**Authors:** Ali Seiphoori[1,4], Xiao-guang Ma[2,5], Paulo E. Arratia[3] and Douglas J. Jerolmack[1,3, *]

**Affiliations:**

[1]Department of Earth & Environmental Science, University of Pennsylvania, Philadelphia, PA 19104, USA.

[2]Department of Physics & Astronomy, University of Pennsylvania, Philadelphia, PA 19104, USA.

[3]Department of Mechanical Engineering & Applied Mechanics, University of Pennsylvania, Philadelphia, PA 19104, USA.

[4]Department of Earth, Atmospheric, & Planetary Sciences, Massachusetts Institute of Technology, Cambridge, MA 02139, USA.

[5]Complex Assemblies of Soft Matter, CNRS-Solvay-UPenn UMI 3254, Bristol, PA 19007, USA.

*Correspondence to: *sediment@sas.upenn.edu*

**Abstract:** When a colloidal suspension is dried, capillary pressure may overwhelm repulsive electrostatic forces, assembling aggregates that are out of thermal equilibrium. This poorly understood process confers cohesive strength to many geological and industrial materials. Here we observe evaporation-driven aggregation of natural and synthesized particulates, probe their stability under rewetting, and measure bonding strength using an atomic force microscope. Cohesion arises at a common length scale (~5 μm), where interparticle attractive forces exceed particle weight. In polydisperse mixtures, smaller particles condense within shrinking capillary bridges to build stabilizing "solid bridges" among larger grains. This dynamic repeats across scales forming remarkably strong, hierarchical clusters, whose cohesion derives from grain size rather than mineralogy. Results may help to understand and control the stability of natural soils and synthetic materials.

**One Sentence Summary:** An evaporating droplet size-segregates particles to make novel fractal aggregates with remarkable strength, and broad implications.

**Main Text:** Understanding the stability and strength of particle assemblies subject to disruptive forces is essential for predicting the macroscopic mechanical behavior of geological, biological, and industrial particulate systems *(1-3)*. Hallmarks of such systems include size polydispersity and non-equilibrium dynamics *(4,5)*. For instance, under favorable conditions, small particles in a polydisperse mixture may bind larger cohesionless grains together by bridging the interparticle space, forming aggregates. This mechanism is manifested in various industrial processes such as the flocculation of polymers *(6,7)*, cement binding of frictional grains *(8,9)*, contact fusion of metal particles during sintering *(10,11)*, and agglomeration of various industrial products including commercial fertilizers and pharmaceutical products *(12,13)*. Evaporation of stable (electrostatically-repulsive) colloidal suspensions can form aggregates *(14)*, gels *(15)* and glasses *(16)*, by driving particles together via capillary forces. In addition to electrostatic and electromagnetic attractions *(17)*, various interfacial interactions – such as chemical bonding *(18)*, capillary adhesion and solute (re-)crystallization *(19,20)* can confer strength to the particle assembly and give rise to an effective cohesion. In the natural environment, soil cohesion determines the erosion susceptibility of landscapes including riverbanks, marshes, hillsides and agricultural fields *(21-23)*. Cohesion is an important factor in geotechnical applications *(24,25)*, and a crucial parameter in transport of contaminants and microorganisms in the environment *(26,27)*. Many natural and industrial materials are composed of particles of various size, shape and surface charge, and are subject to intermittent cycles of wetting and drying. The formation and stability of aggregates under these conditions has not been examined at the particle scale, and cannot be predicted from interfacial electrostatic forces.

**Approach for examining assembly and stability of aggregates**

Here we experimentally investigate the assembly of aggregates formed by evaporating various suspensions, and probe the stability of these aggregates subject to controlled rewetting (Fig. 1). We create an idealized model system composed of silica spheres of different sizes ranging from nanometer to micrometer scale, with well-characterized geometry and surface-charge properties. We then compare results for a range of natural materials including clays (see Materials and Methods and Fig. S1). In each experiment, we first deposit a droplet of particle suspension on a borosilicate coverslip placed on an inverted optical microscope, and let it air-dry under laboratory conditions (Fig. 1A). During evaporation, the suspended particles are subjected to both random and directional forces; the former includes Brownian and interparticle electrostatic forces *(28)*, while the latter includes gravity and drag forces due to Marangoni flow *(29)* and an outward capillary flow (*i.e.*, coffee ring flow *(30)*) that may pull suspended particles to the contact line as evaporation proceeds. During evaporation, the flow-induced forces (*i.e.*, the outward convective flow and the capillary force at air-water interface) are dominant *(30,31)*. As evaporation proceeds, a contact line migrates inward in a stick-slip fashion due to partial pinning by defects on the surfaces of both the substrate and the substrate-bound particles. As a result, the suspended particles are either concentrated at a pinned edge, or dragged toward the center by the retreating motion of the interface (Fig. 1C). Finally, the dried particles form unique patterns on the substrate as the result of a competition between the flow-induced forces and the colloid-substrate interaction during evaporation *(32,33)*.

After evaporation, we fully dry the particle deposits by applying a vacuum, and image sample aggregates using Scanning and Transmission Electron Microscopes (SEM and TEM) to obtain the fine details of their structure. We subsequently mount a microfluidic channel onto the coverslip over the aggregates (Fig. 1B; see Materials and Methods). A pump then flows water through the channel and over the aggregate, where rewetting occurs in three steps: first, the increase in humidity in the channel induces water condensation on surfaces of both the substrate and the particles; second, a transient capillary interface (contact line) migrates across the channel; and third, a steady (fully-saturated) laminar flow develops and applies fluid shear to the aggregates (Fig. 1D, E). We note that the capillary force imposed by the moving contact line depends on the wetting properties of the particle surfaces. For partially wetted surfaces (as shown in the model system presented in Fig.1D), the contact line deforms due to pinning by particles. The resultant capillary force is orders of magnitude larger than the drag and lift forces imparted on the particles by the saturated flow (see Fig. S2). As a result, particle entrainment typically occurs by the moving contact line; most aggregates remaining can withstand the subsequent hydrodynamic forces after inundation (supplementary Movie S1).

**Formation of stabilizing solid bridges**

We first examine the drying and rewetting dynamics of a suspension of bidisperse silica microspheres composed of 20-$\mu m$ and 3-$\mu m$ particles with concentration of 2.0 and 0.2 *wt%*, respectively. We note that the surfaces of the silica particles and the coverslip are all negatively charged in water. Thus particle-particle and particle-substrate interactions in suspension are expected to be repulsive (see Fig. S1). In addition, we pre-treat the coverslip to make it more hydrophilic to encourage the formation of widely dispersed, isolated aggregates (rather than one large aggregate as seen in Fig. 1C; see Methods). As evaporation of the droplet begins, the retreating air-water interface drags smaller particles and condenses some of them within the meniscus formed between larger particles and the substrate or between adjacent particles (Fig. 2A-C). Once all water is evaporated, we find that these small particles form '*solid bridges*' that connect larger particles to the substrate and to each other to make aggregates (Fig. 2C; supplementary Movie S2). The deposit is then subject to rewetting by a transient capillary flow with an average velocity, $u$ = 180 $\mu$m/s and Reynolds number, $\Re$ = 0.03 (Fig. 2D-F). We find that the 20-$\mu$m particles that are stabilized through solid bridging are not transported by the flow (Supplementary Movie S3). In contrast, evaporating a monodisperse suspension of 20-$\mu m$ particles generates a deposit without solid bridges that is easily resuspended and transported by the flow (Fig. 2G-I; Supplementary Movie S4). Our results show that small particles are not only stable, but also tend to stabilize larger ones by bridging interparticle space and binding large particles to the substrate or one to another.

**Hierarchical structure of aggregates**

We observe a consistent and intriguing pattern in the aggregates formed from the bidisperse suspension; small particles surround larger ones in a concentric arrangement. It appears that as the droplet evaporates, it breaks up into smaller regions around the larger particles; the retreating contact line then drags small particles radially inward toward each large particle (Fig. 2C). To further probe this pattern forming process, we examine the evaporation of a polydisperse suspension composed of five different particle sizes: 90, 20, 3, 0.4 and 0.02 $\mu m$ at 5.0, 2.0, 0.2, 0.01, and 0.001 *wt*%, respectively (Fig. 3A-D). A multiscale observation of the deposited aggregates reveals a remarkable self-similar, hierarchical structure of particle aggregates formed by solid bridges. The resulting aggregates are strongly size-segregated, with large particles ringed by smaller particles, and those particles ringed by even smaller particles, and so on. Although classical aggregation is known to make fractal structures *(34-36)*, this hierarchical arrangement of particle sizes seems to be a special consequence of evaporation-induced aggregation. Evaporation and breakup of the droplet drives the cascading assembly of aggregates from large to small pore spaces via capillary pressure, which condenses smaller colloids within the capillary bridges at each scale (Fig. 3E). We find that the aggregates formed by polydisperse silica spheres are stable when subject to rewetting.

    We now consider the relevant physical parameters that allow the assembly and stability of aggregates formed by evaporation. We begin by calculating the pair-wise potential energy between particles by incorporating double-layer repulsion and van der Waals attraction potentials

(as in DLVO theory), as well as the surface hydration potential (Supplementary Information). We estimate energy as a function of the surface-to-surface distance, *x*, between two approaching $D_1 = 20$ *nm* silica particles in water during evaporation (Fig. 3F), where $D_1$ is the smallest particle size in the model system, as described in Fig. 3. We find that the maximum energy of the capillary bridge between these particles under evaporation ($\gamma D_1^2 \sim 10^3 k_B T$) is much larger than the repulsive barrier, and should easily push particles within the range of van der Waals attraction *(37,38)*. Furthermore, full dehydration of the surfaces (*e.g*, by applying vacuum as in our experiments) further enhances the van der Waals attraction (Fig. 3F). Thus, evaporation can create strong aggregates, and strong bonds to the substrate, by collecting a large number of small particles of size $D_1$ to the contact points of larger particles of size $D_2$. This way, particles of size $D_i$ can bridge and bind the neighboring particles of size $D_{i+1}$ to form a multiscale aggregate system.

**Solid bridging in naturally-occurring particles**

To further test the generality of stabilization by solid bridging, we investigate suspensions of two structurally distinct, naturally-occurring clay types: illite, the most common clay mineral with a 2:1 (TOT phyllosilicate) structure; and kaolinite, a widely used industrial clay with a simple 1:1 (TO phyllosilicate) structure *(39)*. Despite having more complex shapes and surface-charge properties (see Fig. S1), we identify a similar behavior in clay particles to the idealized silica spheres. Consider a polydisperse suspension of illite particles ranging from 0.1 *μm* < *D* < 100 *μm*. We find that evaporation forms solid bridges and strong aggregates that are stable to rewetting (Fig. 4A-C; Supplementary Movie S5). Sieving the illite clay to remove particles with *D* < 5 *μm*, however, leads to aggregates that disintegrate when rewetted (Fig. 4D-F; Supplementary Movie S6) suggesting that polydispersity is critical for solid bridge formation.

Kaolinite clay particles exhibit similar behavior to illite (Supplementary Movies S7 and S8). These results demonstrate that solid bridging induces an effective cohesion in polydisperse aggregates, and that this effect is insensitive to material properties.

**Determining the strength conferred by solid bridges**

To quantify the strength of bonding between the particles and the substrate, we use an atomic force microscope (AFM) to determine the detachment (pull-off) force and energy associated with 20-$\mu$m silica spheres (Fig.5; Methods). We first deposit a monodisperse suspension of 20-$\mu$m particles on a coverslip, where solid bridging is not expected. Images indicate, however, that nanoparticulate contamination composed of amorphous silica is nearly unavoidable; this is expected to confer some strength to the particles (see Fig. S3). The AFM force-displacement curves ($F_s$-$z$) initially have a positive force associated with the compression of the cantilever onto the particle (Fig. 5); as the cantilever is lifted, $F_s$ grows increasingly negative until detachment occurs and it abruptly drops to zero. The maximum (pull-off) force measured using this technique for the 20-$\mu$m particles is of order $10^{-8}N$. By summing the area under the force-displacement curve, we find that the energy associated with detaching the particle from the substrate is of order $10^5 k_B T$ (Fig. 5), which is much larger than measured adhesion force between the AFM apex (without epoxy; see Materials and Methods) and the particle surface. This shows how nanoparticle contaminants can confer bonding strength through the solid bridging mechanism, even in monodisperse systems.

To examine the strength conferred by solid bridges, we perform a similar AFM experiment with 20-$\mu$m particles deposited from a polydisperse suspension that also contains grains of size 3, 0.4 and 0.02 $\mu$m. The force-displacement curve in this case displays multiple stick-slip events (Fig. 5A), that are likely due to the breaking of individual or groups of bonds associated with solid bridges. The solid bridges in this polydisperse deposit increase the detachment energy by approximately 20 times (Fig. 5A). We find similar force-displacement behavior for the bidisperse system of 20-$\mu$m and 3-$\mu$m particles studied above (see Fig. S4). This observation suggests that the number of breaking events are associated with a debinding mechanism at the coordination points, where the bonds between the large particle and the surrounding smaller particles are developed. We note that the measured pull-off forces and energies using the AFM may not be reflective of conditions associated with detachment by water; hydration due to humidity precedes the wetting front (Supplementary Movie S9), and could weaken inter-particle bonds *(40)* relative to the conditions of the AFM. Nevertheless, AFM results provide a useful relative measure of the stability conferred on large particles by solid bridging of smaller particles.

**Discussion**

It is believed that clays aggregate due to physicochemical surface properties that induce electrostatic attraction in solution *(41-43)*. While surface properties such as charge type and density influence particle interactions in suspension, our results show how interfacial capillary force can dominate over electrostatic effects. During evaporation, particles tend to be concentrated at the contact points, where attractive bonds develop. Those individual bonds may be weak (e.g., short range van der Waals type), but the sum of attractive forces and the resultant cohesion can be remarkably large, especially for systems with a large number of contacts. Such behavior is facilitated by the presence of small particles, that effectively increase the specific surface area of the system.

True cohesion in natural soils is typically attributed to the presence of clay minerals, or other cementing agents such as carbonates, iron oxides, and salts *(39,44)*. We have shown that a typical clay material effectively loses its cohesion when small particles are absent, while silica particles (≤ 5 *μm*) with simple interparticle contact properties form cohesive aggregates (see Fig. S5). Indeed, for a range of materials we observe a transition from cohesionless to cohesive aggregates at a common length scale, where interparticle forces exceed the particle weight. Thus, attributing the cohesion in natural soils to material properties (*e.g.*, by introduction of clay minerals) might obscure the contribution made by the particle size. Soils are composed of polydisperse particles subject to cycles of evaporation and rewetting. It is possible that the formation and strength of soil aggregates has more to do with size than with material composition. The wetting behavior of soils is of fundamental importance for soil collapse *(45,46)*, creep and liquefaction in landslides *(24,47)*, and slaking erosion that results from disintegration of aggregates *(23,25)*. Our work provides a new physical basis, and a new set of tools, for understanding the rewetting behavior of soil aggregates. More broadly, our results open the path to studying the effect of particle size polydispersity and transient hydrodynamic forces on the multiscale mechanical behavior of particulate assemblies ranging from industrial to complex living systems. In addition, hierarchical assembly of aggregates by evaporation may provide a novel method for creating thin films with desired mechanical properties *(48,49)*.

**References and Notes:**

1.  Chiti, F., Stefani, M., Taddei, N., Ramponi, G. & Dobson, C. M. Rationalization of the effects of mutations on peptide and protein aggregation rates. *Nature* **424**, 805–808 (2003).
2.  Mueller, S. B. *et al*. Stability of volcanic ash aggregates and break-up processes. *Sci. Rep.* **7**, 7440 (2017).
3.  Zhang, H. *et al.* Stable colloids in molten inorganic salts. *Nature* **542**, 328–331 (2017).
4.  Anderson, V. J., & Lekkerkerker, H. N. Insights into phase transition kinetics from colloid science. *Nature* **416**, 811–815 (2002).
5.  Xia, Y. *et al.* Self-assembly of self-limiting monodisperse supraparticles from polydisperse nanoparticles. *Nat. Nanotechnol.* **6**, 580–587 (2011).
6.  Swenson, J., Smalley, M. V & Hatharasinghe, H. L. M. Mechanism and strength of polymer bridging flocculation. *Phys. Rev. Lett.* **81**, 5840–5843 (1998).
7.  Biggs, S., Habgood, M. & Jameson, G. J. Aggregate structures formed via a bridging flocculation mechanism. *Chem. Eng. J.* **80**, 13–22 (2000).
8.  Fonseca, P. C. & Jennings, H. M. The effect of drying on early-age morphology of C-S-H as observed in environmental SEM. *Cem. Concr. Res.* **40**, 1673–1680 (2010).
9.  Zhou, T., Ioannidou, K., Ulm, F. J., Bazant, M. Z., & Pellenq, R. M. Multiscale poromechanics of wet cement paste. *Proc. Natl. Acad. Sci. U. S. A.* **116**, 10652–10657 (2019).
10. Wanke, S. E. & Flynn, P. C. The sintering of supported metal catalysts. *Catal. Rev.* **12**, 93–135 (1975).
11. Dynys, F. & Halloran, J. W. Influence of aggregates on sintering. *J. Am. Ceram. Soc.* **67**, 596–601 (1984).
12. Bika, D., Tardos, G. I., Panmai, S., Farber, L., & Michaels, J. Strength and morphology of solid bridges in dry granules of pharmaceutical powders. *Powder Technol*. **150**, 104–116 (2005).
13. Avilés-Avilés, C., Dumoulin, E. & Turchiuli, C. Fluidised bed agglomeration of particles with different glass transition temperatures. *Powder Technol.* **270**, 445–452 (2015).
14. Liu, M., Guo, B., Du, M. & Jia, D. Drying induced aggregation of halloysite nanotubes in polyvinyl alcohol/halloysite nanotubes solution and its effect on properties of composite film. *Appl. Phys. A* **88**, 391–395 (2007).
15. Pauchard, L., Parisse, F. & Allain, C. Influence of salt content on crack patterns formed through colloidal suspension desiccation. *Phys. Rev. E* **59**, 3737–3740 (1999).
16. Haw, M. D., Gillie, M. & Poon, W. C. K. Effects of phase behavior on the drying of colloidal suspensions. *Langmuir* **18**, 1626–1633 (2002).
17. Lu, P. J. *et al.* Gelation of particles with short-range attraction. *Nature* **453**, 499 (2008).
18. Roy, D. M. New strong cement materials: chemically bonded ceramics. *Science* **235**, 651–658 (1987).

19. Soulié, F., El Youssoufi, M. S., Delenne, J. Y., Voivret, C., & Saix, C. Effect of the crystallization of a solute on the cohesion in granular materials. *Powder technol.* **175**, 43–47 (2007).
20. Jones, R., Pollock, H. M., Cleaver, J. A., & Hodges, C. S. (2002). Adhesion forces between glass and silicon surfaces in air studied by AFM: Effects of relative humidity, particle size, roughness, and surface treatment. *Langmuir*, 18(21), 8045-8055.
21. Smalley, I. J. Cohesion of soil particles and the intrinsic resistance of simple soil systems to wind erosion. *J. Soil Sci.* **21**, 154–161 (1970).
22. Howard, A. D. & McLane, C. F. Erosion of cohesionless sediment by groundwater seepage. *Water Resour. Res.* **24**, 1659–1674 (1988).
23. Barthes, B. & Roose, E. Aggregate stability as an indicator of soil susceptibility to runoff and erosion; validation at several levels. *Catena* **47**, 133–149 (2002).
24. Ishihara, K. Liquefaction and flow failure during earthquakes. *Géotechnique* **43**, 351–451 (1993).
25. Li, M. *et al*. Modelling soil detachment by overland flow for the soil in the Tibet Plateau of China. *Sci. Rep.* **9**, 8063 (2019).
26. McCarthy, J. & Zachara, J. Subsurface transport of contaminants. *Environ. Sci. Technol.* **23**, 496–502 (1989).
27. Garrett, T. R., Bhakoo, M., & Zhang, Z. Bacterial adhesion and biofilms on surfaces. *Prog. Nat. Sci.*, **18**, 1049–1056 (2008).
28. Allain, C. & Limat, L. Regular Patterns of Cracks Formed by Directional Drying of a Colloidal Suspension. *Phys. Rev. Lett.* **74**, 2981–2984 (1995).
29. Maillard, M., Motte, L., Ngo, A. T. & Pileni, M. P. Rings and hexagons made of nanocrystals: A Marangoni effect. *J. Phys. Chem. B* **104**, 11871–11877 (2000).
30. Deegan, R. D. *et al.* Capillary flow as the cause of ring stains from dried liquid drops. *Nature* **389**, 827–829 (1997).
31. Weon, B. M. & Je, J. H. Capillary force repels coffee-ring effect. *Phys. Rev. E* **82**, 015305 (2010).
32. Nadkarni, G. D. & Garoff, S. An investigation of microscopic aspects of contact angle hysteresis: Pinning of the contact line on a single defect. *Europhys. Lett.* **20**, 523–528 (1992).
33. Deegan, R. D. Pattern formation in drying drops. *Phys. Rev. E* **61**, 475–485 (2000).
34. Schaefer, D. W., Martin, J. E., Wiltzius, P. & Cannell, D. S. Fractal Geometry of Colloidal Aggregates. *Phys. Rev. Lett.* **52**, 2371–2374 (1984).
35. Lin, M. Y. *et al.* Universality in colloid aggregation. *Nature* **339**, 360–362 (1989).
36. Kranenburg, C. The fractal structure of cohesive sediment aggregates. *Estuar. Coast. Shelf. Sci.*, **39**, 451–460 (1994).
37. Israelachvili, J. N. *Intermolecular and Surface Forces*. 3rd edn. (Academic press, Amsterdam, 2011).

38. Rabinovich, Y. I., Esayanur, M. S. & Moudgil, B. M. Capillary forces between two spheres with a fixed volume liquid bridge: theory and experiment. *Langmuir* **21**, 10992–10997 (2005).

39. Mitchell, J. K. & Soga, K. *Fundamentals of Soil Behavior*. 3rd edn. (John Wiley & Sons, Hoboken, 2005).

40. Israelachvili, J. & Wennerström, H. Role of hydration and water structure in biological and colloidal interactions. *Nature* **379**, 219–225 (1996).

41. Lewis, J. A. Colloidal processing of ceramics. *J. Am. Ceram. Soc.* **83**, 2341–2359 (2000).

42. Berka, M. & Rice, J. A. Absolute aggregation rate constants in aggregation of kaolinite measured by simultaneous static and dynamic light scattering. *Langmuir* **20**, 6152–6157 (2004).

43. Ruzicka, B. *et al*. Observation of empty liquids and equilibrium gels in a colloidal clay. *Nat. Mater.* **10**, 56–60 (2011).

44. Winterwerp, J. C., & Van Kesteren, W. G. *Introduction to the Physics of Cohesive Sediment Dynamics in the Marine Environment*. 1st edn. (Elsevier. Amsterdam, 2004).

45. Barden, L., McGown, A. & Collins, K. The collapse mechanism in partly saturated soil. *Eng. Geol.* **7**, 49–60 (1973).

46. Smalley, I. J., Ross, C. W. & Whitton, J. S. Clays from New Zealand support the inactive particle theory of soil sensitivity. *Nature* **288**, 576–577 (1980).

47. Gabet, E. J., & Mudd, S. M. The mobilization of debris flows from shallow landslides. *Geomorphology* **74**, 207–218 (2006).

48. Brinker, C. J., Lu, Y., Sellinger, A. & Fan, H. Evaporation-induced self-assembly: nanostructures made easy. *Adv. Mater.* **11**, 579–585 (1999).

49. Rabani, E., Reichman, D. R., Geissler, P. L., & Brus, L. E. Drying-mediated self-assembly of nanoparticles. *Nature* **426**, 271–274 (2003).

50. Xia, Y. & Whitesides, G. M. Soft lithography. *Angew. Chemie Int. Ed.* **37**, 550–575 (1998).

51. Lameiras, F. S., Souza, A. L. de, Melo, V. A. R. de, Nunes, E. H. M. & Braga, I. D. Measurement of the zeta potential of planar surfaces with a rotating disk. *Mater. Res.* **11**, 217–219 (2008).

52. Wiberg, P. L. & Smith, J. D. Calculations of the critical shear stress for motion of uniform and heterogeneous sediments. *Water Resour. Res.* **23**, 1471–1480 (1987).

53. Derjaguin, B. v & Landau, L. Theory of the stability of strongly charged lyophobic sols and of the adhesion of strongly charged particles in solutions of electrolytes. *Prog. Surf. Sci.* **43**, 30–59 (1993).

54. Verwey, E. J. W. Theory of the stability of lyophobic colloids. *J. Phys. Chem.* **51**, 631–636 (1947).

55. Hamaker, H. C. The London—van der Waals attraction between spherical particles. *physica* **4**, 1058–1072 (1937).

56. Bergström, L. Hamaker constants of inorganic materials. *Adv. Colloid Interface Sci.* **70**, 125–169 (1997).

57. Hogg, R., Healy, T. W. & Fuerstenau, D. W. Mutual coagulation of colloidal dispersions. *Trans. Faraday Soc.* **62**, 1638–1651 (1966).

58. Van Oss, C. J., Giese, R. F. & Costanzo, P. M. DLVO and non-DLVO interactions in hectorite. *Clays Clay Min.* **38**, 151–159 (1990).

59. Van Oss, C. J. *Interfacial Forces in Aqueous Media*. 2nd edn. (CRC press, New York 2006).

60. Pashley, R. M. Hydration forces between mica surfaces in aqueous electrolyte solutions. *J. Colloid Interface Sci.* **80**, 153–162 (1981).

61. Pashley, R. M. DLVO and hydration forces between mica surfaces in Li+, Na+, K+, and Cs+ electrolyte solutions: A correlation of double-layer and hydration forces with surface cation exchange properties. *J. Colloid Interface Sci.* **83**, 531–546 (1981).

62. Hunter, R. J. *Foundations of Colloid Science*. 2nd edn. (Oxford university press, Oxford, 2001).

**Acknowledgments:**

We thank M. Brukman, H. Khare, and R. Vigliaturo for useful discussions and comments.

**Funding**

Authors would like to acknowledge financial support from the U.S. National Institute of Environmental Health Sciences (NIEHS), Grant Number P42ES02372, and the U.S. Army Research Office (ARO), Grant Number 569074. P.E.A. acknowledges partial support from the Materials Research Science and Engineering Center (MRSEC), DMR-1720530.

**Author Contributions**

A.S. and D.J.J initiated the research. A.S., D.J.J and P.E.A. designed the research. A.S. performed the experiments, and analyzed the data. All authors interpreted the results and contributed to writing the manuscript.

**Competing Interests**

The authors declare no competing interests.

Correspondence and requests for materials should be addressed to A.S. or D.J.J. at sediment@sas.upenn.edu.

**Data and Material Availability**

Experimental data and the code from this study are publicly available at https://github.com/seiphoori/Solidbridging

**Supplementary Materials:**

Materials and Methods

Supplementary Text

Figs. S1 to S5

Movies S1 to S10

References (*50-62*)

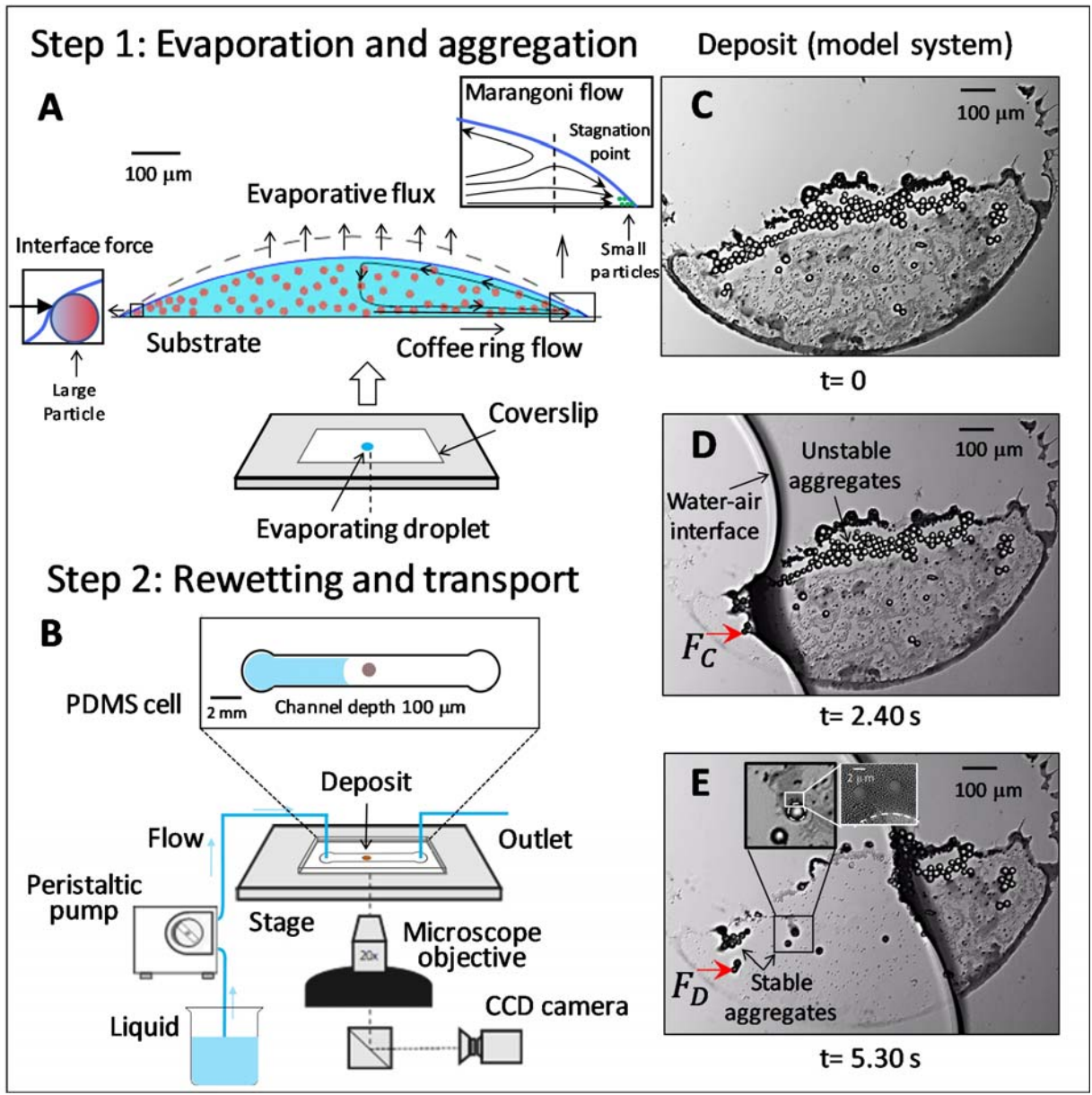

**Fig. 1. Approach for examining experimental assembly and stability of aggregates.** (**A** and **B**) Experimental design. (**C**), Evaporation pattern of a polydisperse system composed of silica spheres. (**D**), Transport mechanism due to a transient capillary interface (contact line) migrating across the channel, imposing a capillary force on particles, $F_C$. (**E**), Transport mechanism due to a laminar flow regime, imposing a drag force on particles $F_D$.

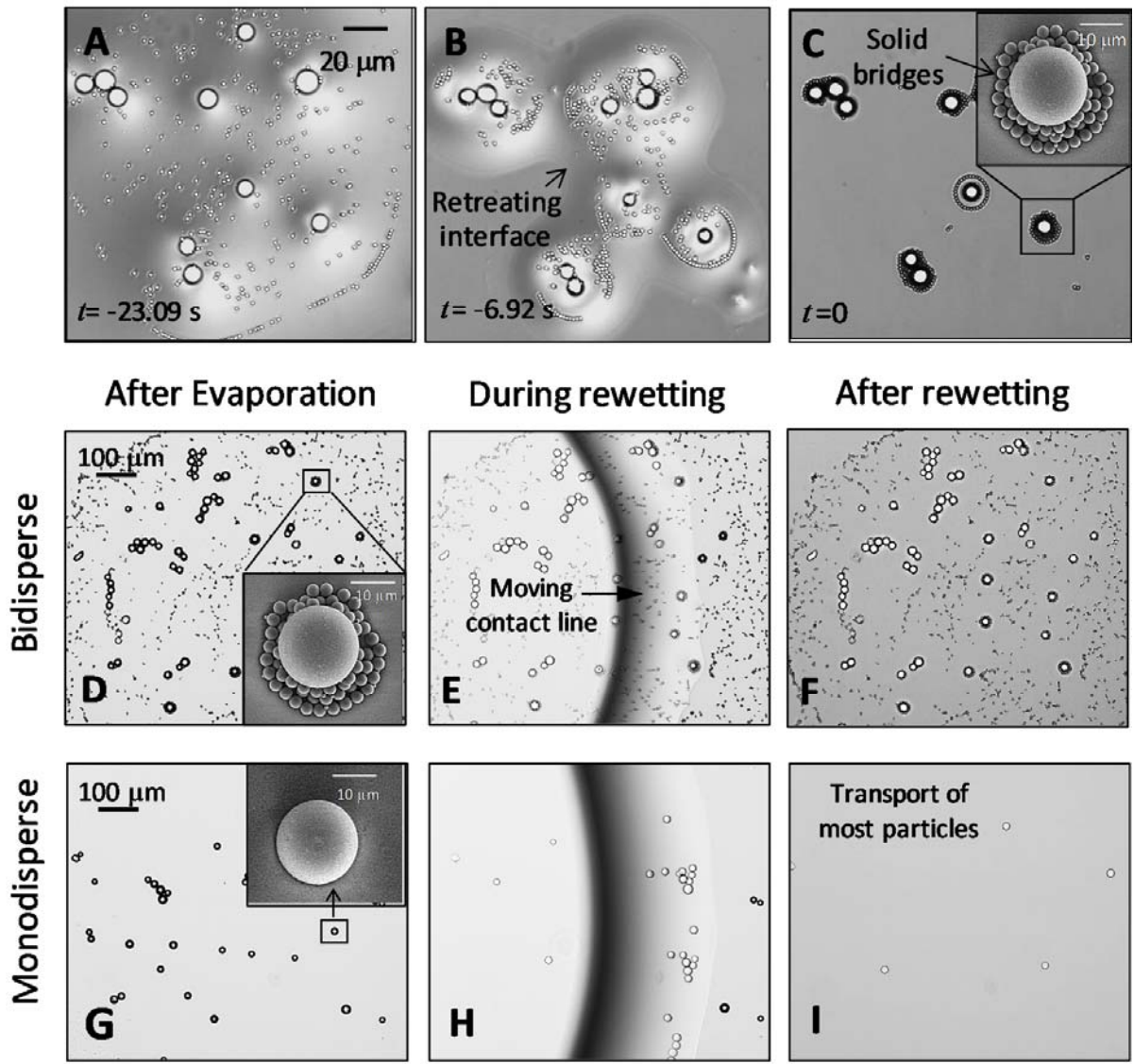

**Fig. 2. Contribution of solid bridges to stabilizing the evaporation-induced aggregates.** (**A** and **B**) Evaporation from a bidisperse **suspension** composed of 20 and 3-$\mu m$ silica spheres, where smaller particles condense within the diminishing capillary bridges. (**C**) Formation of solid bridges between larger particles or between particles and the substrate. (**D** to **F**) Stability of aggregates due to the presence of solid bridges. (**G** to **I**) Transport of 20-$\mu m$ particles in the absence of smaller particles and the associated solid bridges.

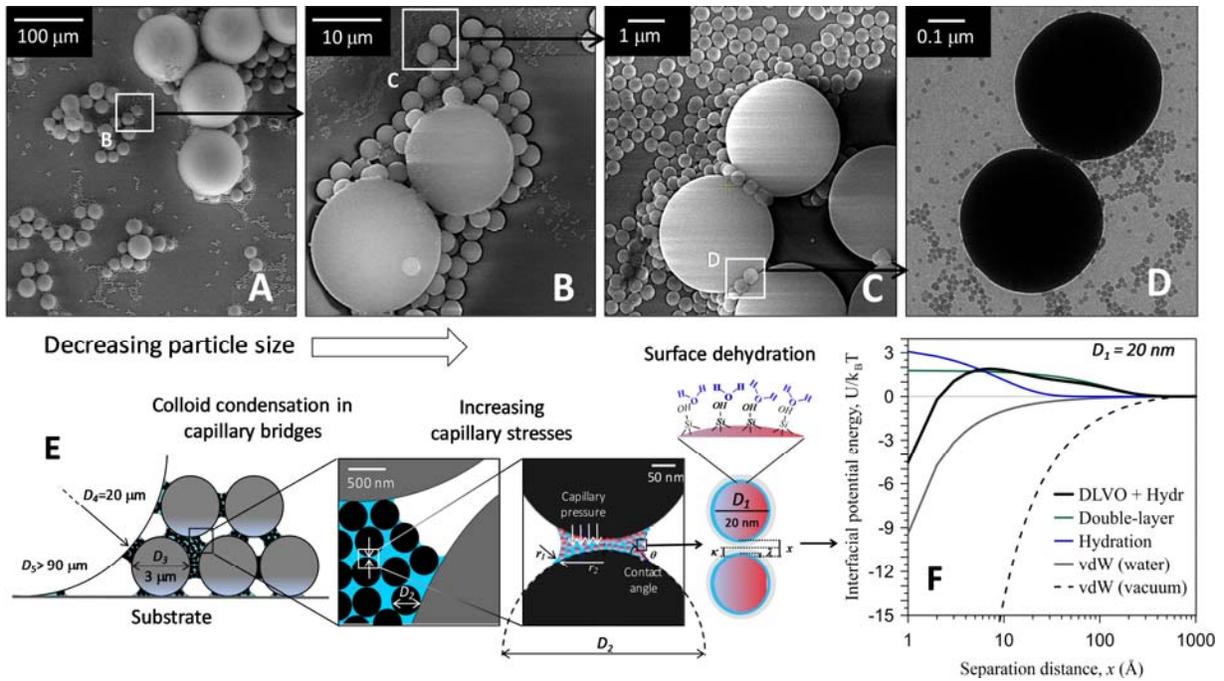

**Fig. 3. Hierarchical structure of aggregates.** (**A** to **D**) Multi-scale observation of a polydisperse colloidal system composed of silica spheres with particle sizes of 90, 20, 3, and 0.4 *μm* along with silica nanoparticles of 20-*nm* size. (**E**) Capillary force drives particles together by overcoming the interparticle repulsion. Further dehydration of the particle surface is energetically favorable resulting in an overall van der Waals attraction between particles. (**F**) Interfacial potential energy functions between two particles of $D_1 = 20$ *nm* size, calculated from theory (see Supplementary Information).

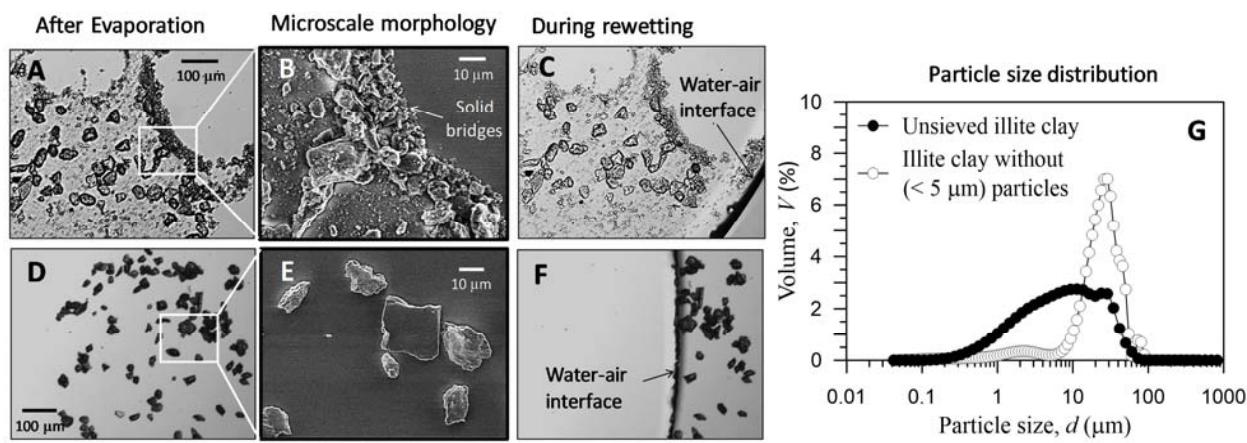

**Fig. 4. Stability of aggregates of illite clay particles through formation of solid bridges**. (**A**) Aggregates formed by unsieved, polydisperse illite particles. (**B**) Morphology of the solid bridges at microscopic scale. (**C**) Aggregates are stable when subject to rewetting, due to solid bridges. (**D**) Aggregates of same illite clay modified by sieving out particles < 5 $\mu m$. (**E**) Morphology of illite particle deposits. (**F**) Removal of small particles makes aggregates unstable to rewetting. (**G**) Particle size distributions of unsieved and sieved illite particles suspensions.

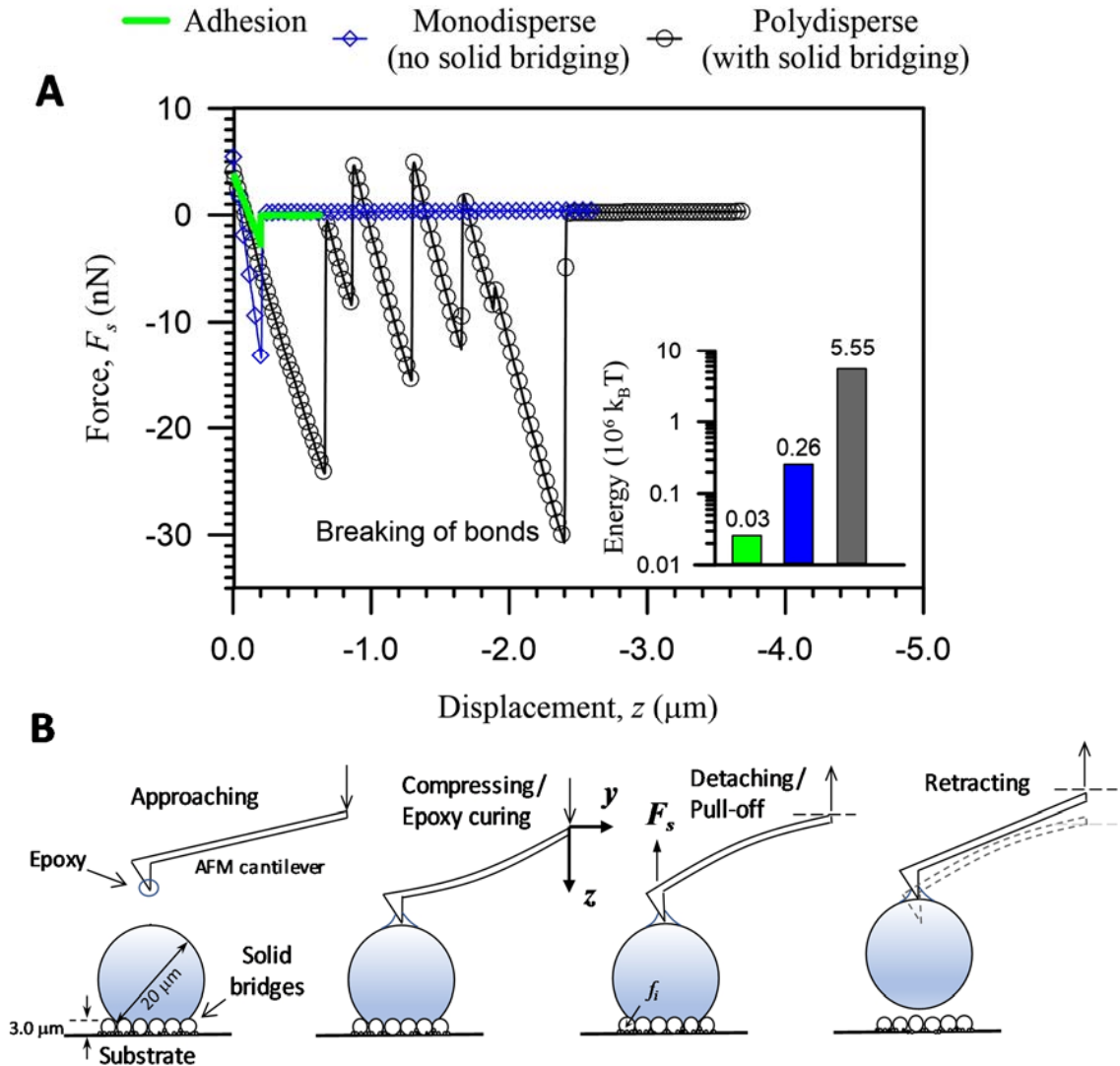

**Fig. 5. Pull-off force measured with Atomic Force Microscope (AFM) using contact mode technique**. (**A**) Force-displacement ($F_s$-$z$) and detachment energy (inset) of a single 20-µm silica particle without and with solid bridges after evaporation from a monodisperse and polydisperse suspension, respectively. The adhesion force between the apex of the AFM cantilever (without epoxy) and the surface of same particles is also shown for reference. (**B**) Modifying the AFM contact mode procedure for pull-off experiments. $F_s$ is the measured force associated with breaking of individual bonds or groups of bonds at the contact points ($f_i$).

# Supplementary Materials for

## Formation of stable aggregates by fluid-assembled solid bridges

Ali Seiphoori*, Xiao-guang Ma, Paulo E. Arratia and Douglas J. Jerolmack*

Correspondence to: sediment@sas.upenn.edu

**This PDF file includes:**

    Materials and Methods
    Supplementary Text
    Figs. S1 to S5
    Captions for Movies S1 to S10

**Other Supplementary Materials for this manuscript include the following:**

    Movies S1 to S10

**Materials and Methods**

**Particle properties and preparation of suspensions.** Various silicate-based colloids were selected to encompass a wide range of relative charge, shape and size heterogeneity. Aqueous suspensions of silica microspheres (Corpuscular Inc., NY, USA) with mean diameter of 20, 3, 0.4 and 0.02 $\mu m$ (particle density, $\rho_s$ = 2.65 $g/cm^3$) were diluted in deionized water (Milli-Q Reagent Water System, Millipore, MA, USA) and mixed to form an idealized polydisperse model system. Mined phyllosilicate clay minerals including illite (pulverized Cambrian shale from Silver Hill formation) and kaolinite (Washington County formation, GA, USA) were used to represent natural soil particles. The particle density of illite and kaolinite was measured as $\rho i$ = 2.77 and $\rho k$ = 2.65 $g/cm^3$, respectively. The BET surface area of the illite and kaolinite particles was measured as 10.98 $m^2/g$, and 9.09 $m^2/g$ using a Micromeritics 3Flex surface area analyzer (Micromeritics Instrument Co., GA, USA). Dry clay particles were initially dispersed in deionized water and allowed to rest for 24 hours. The suspension was then sonicated and passed through a 90 $\mu m$ sieve (#170 ASTM-E11, ASTM, PA, USA) to filter out large aggregates. The particle size distribution of the clay suspension was computed using a laser diffraction analyzer equipped with an ultrasonic dispersing module (Beckman Coulter, Inc., GA, USA). To produce a monodisperse clay suspension, small particles (< 5 $\mu m$) were removed through multiple wet-sieving (5-$\mu m$ mesh screen (316L), Utah Biodiesel Supply, UT, USA) and ultrasonic washing of the particles submerged in deionized water. After each attempt, the particle size distribution of the suspension was measured, and samples were examined using an optical microscope to ensure the elimination of small particles. The electrophoretic mobility of the colloidal suspension of silica microspheres (3-$\mu m$ size) and clay particles (size <5 $\mu m$) were measured at various pH values using a ZetaPALS instrument (Brookhaven, NY, USA), on 15-20 ppm particle suspensions at temperature 25.0 ± 0.5 ◦C. The pH values were adjusted by dropwise addition of HCl or KOH solutions. Given the polydispersity of the shape and size of the clay particles, the zeta potentials were estimated using Helmholtz-Smoluchowski equation. The average zeta potential of silica spheres, illite particles, and kaolinite particles were measured as –54.9, –27.7, and –32.6 *mV* for pH = 7.0± 0.5, where the evaporation and rewetting experiments were conducted.

**Evaporation and rewetting experiments.** The colloidal droplets (approximately 0.1 *μL*) were initially placed on a borosilicate glass coverslip (Fisher Scientific, NH, USA; thickness varying from 0.13 to 0.17 *mm*). The droplet was then air-dried due to evaporation under laboratory conditions (RH =0.50± 0.05, T=22◦C). The coverslip glass has an average RMS roughness of 13 ± 3.5 Å as measured by AFM technique (Icon, Bruker Co., MA, USA). Pre-cleaning the coverslip by dipping in acetone for 5 min, rinsing with isopropyl alcohol and deionized water, and then drying using compressed nitrogen gas, resulted in a semi-hydrophilic surface (contact angle 30.0± 2.5º) that tended to lead to clumped aggregates (as in Fig.1). To encourage the formation of widely dispersed, isolated aggregates after evaporation, the coverslip surface was pre-treated by using $O_2$-plasma (SCE-108 Barrel Asher, Anatech USA, CA, USA) under a radio frequency (RF) power of 50 *W* for a duration of 15 *s* under chamber pressure of 300 *m Torr*. This treatment provides a more hydrophilic surface with a small contact angle for the water film on the glass (contact angle 8.0± 2.5º). By creating isolated aggregates, we are able to perform AFM pull-off experiments on a single 20-*μm* particle bound to the substrate through solid bridges formed by smaller particles. The deposit was then placed in a vacuum chamber (500 *m Torr*), and after another $O_2$-plasma treatment (50 *W* for 15 s), a microfluidic device was mounted on the coverslip so that the deposit was aligned within a rectangular channel with dimensions of 100 $\mu m$ height, 2 mm width, and 20 mm length. The microfluidic device was made of polydimethylsiloxane (PDMS) using standard soft-lithography methods *(50)*. The fluid was pumped through the channel via a positive displacement peristaltic pump (Minipuls 3, Gilson, WI, USA) under a non-slip boundary condition. The channel was then subject to a rehydration path through a steady and laminar flow.

**Multiscale observation.** The microfluidic device was positioned on an inverted microscope, Elipse Ti-E, (Nikon Instrument Inc., NY, USA), with a 20*x* objective (N.A.= 0.75, resolution=0.37 *µm*), where the images were acquired using an Andor iXon3 EMCCD camera by NIS Elements software (Nikon Instrument Inc., NY, USA). SEM images were acquired using an Environmental Scanning Electron Microscope, FEI Quanta 600 FEG, (Thermo Fisher Scientific, MA, USA) operating at 10–30 *keV* equipped with an electron backscatter diffraction (EBSD) analyzer to provide the energy-dispersive X-ray spectroscopy (EDXS) for chemical phase analysis. TEM images were acquired using a Field Emission Scanning Transmission Electron Microscope, JEOL F200, (JEOL, MA, USA) operating at 120 *keV*. Selected Area (Electron) Diffraction (SAED) technique was employed to obtain information on the crystal structure of the solid bridges.

**AFM measurements.** The detachment (pull-off) force was measured by modifying the contact mode of an AFM Asylum MFP-3D unit (Asylum Research, CA, USA) mounted on top of the same inverted microscope system (Elipse Ti-E), and the force spectroscopy data were obtained using the IGOR PRO program (WaveMetrics, OR, USA). Soft contact mode cantilevers made of silicon nitride (NanoAndMore Co., CA, USA), 250-350 *µm* long with spring constant of 0.03-0.08 *N/m*, and Al backside coating, were used to cover the range of measured forces. Prior to each experiment, the exact spring constant value of each cantilever was determined using the thermal calibration method. Devcon 5-Minute Epoxy (Devcon Co., MA, USA) was then prepared according to the manufacturer's instructions directly on a coverslip strip and flattened to form a thin film. The apex of the cantilever was approached to the film until contact with the epoxy surface was observed. The cantilever was then pulled up immediately to prevent an excessive amount of epoxy due to capillary action. After depositing a small bead of epoxy, the apex of the cantilever was moved on top and pushed onto the target particle. The epoxy was then allowed to cure for 15 *min* and solidify. The experiment was performed by lifting the cantilever, while deflection data were recorded. The scan rate (loading/unloading rate) was 0.50 *Hz* in all experiments. The setpoint, or zero deflection position in the non-contact regime was set to the zero-voltage position for all force curves.

**Supplementary Text**

**I. Forces acting on particles during rewetting.** During the rewetting of the deposits, the particle surfaces are initially hydrated through the vapor phase (Supplementary Movies 9 and 10), while the capillary interface (contact line) migrates through the channel (Fig. S2). When the contact line reaches particles, it deforms and a cusp forms (Fig. S2a). In equilibrium, the pinning force at the cusp location can be written as:

$$F_{pin} = F_C + F_D , \qquad [S1]$$

where $F_C$ and $F_D$ are the capillary and drag forces imparted on the particle. The capillary force is a function of the wettability of particle surfaces and can be estimated from the Laplace equation *(36)*:

$$F_C = -\gamma \left(\frac{1}{r_1} + \frac{1}{r_2}\right) A_w = -\gamma \frac{A_w}{r_k}, \qquad [S2]$$

where $\gamma$ is the surface tension of water (73 *mN/m*), $r_1$ and $r_2$ are the radii of the water meniscus shaping the cusp around the pinning area, $r_k$ is the Kelvin radius, and $A_w$ is the wetted area of the particle surface (i.e., embracing area). For the geometry shown in the insert of Fig. S2a, the maximum capillary force acting on the particle with diameter $D = 20\ \mu m$, is estimated as 240 $nN$ (order of $10^{-7}\ N$). For wetting surfaces (strong hydrophilicity) with negligible or no pinning, the moving contact line has a smooth geometry (*e.g.*, a line segment) and the capillary force decreases due to an increase in the Kelvin radius. The order of magnitude of the other relevant forces are $\approx 100\ pN$ for particle's weight, and $\approx 30\ pN$ for drag force, $F_D$, under the water flow with a steady state velocity. The drag force can be written as *(52)*:

$$F_D = \frac{1}{2}\rho C_D <u^2(z)> A_x , \qquad [S3]$$

where $u(z)$ is the flow velocity over the particle cross section, $C_D$ is the drag coefficient, and $A_x$ is the cross-sectional area of the particle. We assume a steady, laminar flow with a parabolic flow profile. The drag coefficient can be determined from the experimental relationship for a sphere with diameter $D$ as a function of the particle Reynolds number, $\Re = u > D/\vartheta$, where $\vartheta$ is the kinematic fluid velocity. In a creeping flow regime (Stokes regime) with small Reynolds numbers ($\Re < 0.5$), $C_D = \frac{24}{R}$. The lift force, $F_L$, is negligible ($\approx 100\ fN$), and can be determined in a similar manner to drag force *(52)*:

$$F_L = \frac{1}{2}\rho C_L(u_T^2 - u_B^2)A_x, \qquad [S4]$$

where $C_L$ is the lift coefficient, and $u_T$ and $u_b$ are the flow velocities at the top and bottom of the particle, respectively. We assume a lift coefficient, $C_L = 0.2$ throughout our calculations.

**II. The interfacial potential energy functions.** The interaction of two identical colloidal spheres approaching each other was evaluated based on the Derjaguin-Landau-Verwey-Overbeek (DLVO) theory by superposing the van der Waals and electrostatic double layer potential energies *(53,54)*. The total potential energy, *U*, can be evaluated as a function of the particles surface-to-surface separation distance, *x*:

$$U(x) = U_{vdw}(x) + U_{edl}(x), \tag{S5}$$

The van der Waals potential, $U_{vdw}$, was estimated using the following equation *(55)*:

$$U_{vdw}(x) = \frac{-A_{132}}{12}\left\{\frac{1}{r^2+2r} + \frac{1}{r^2+2r+1} + 2\ln\frac{r^2+2r}{r^2+2r+1}\right\}, \tag{S6}$$

where $r = \frac{2x}{D}$, *D* is the spherical particle diameter, and $A_{132}$ is the Hamaker constant for solid 1 and 2 in medium type 3:

$$A_{132} = \left(\sqrt{A_{11}} - \sqrt{A_{33}}\right)\left(\sqrt{A_{22}} - \sqrt{A_{33}}\right), \tag{S7}$$

where $A_{11}$, $A_{22}$, and $A_{33}$ are the Hamaker constants for each component. For identical solids, the $A_{132}$ sign is always positive and corresponds to an attractive van der Waals interaction. The value of $A_{132}$ was calculated to be $4.6 \times 10^{-21} J$ for the silica-water-silica system *(56)*, with $A_{11} = A_{22} = 6.5 \times 10^{-20} J$ for the amorphous silica, and $A_{33} = 3.5 \times 10^{-20} J$ for water. For a silica-vacuum-silica system, the Hamaker constant can be calculated as $A_{132} = 6.5 \times 10^{-20} J$.

The electrostatic double layer potential between two identical spherical particles, $U_{edl}$, is always repulsive and can be expressed as *(57)*:

$$U_{edl}(x) = \epsilon \psi_0^2 \frac{D}{4} \ln(1 + e^{-\kappa x}), \tag{S8}$$

where $\epsilon$ is the permittivity of the surrounding medium, $\epsilon = \epsilon_w \epsilon_0$, where $\epsilon_w$ is the relative permittivity or dielectric constant of the surrounding medium (78.4 for water in our experiments), $\epsilon_0 = 8.854 \times 10^{-12} \, F/m$ is the vacuum permittivity, and $\psi_0$ is the surface potential of the colloid. The inverse Debye length (*i.e.,* reciprocal double layer thickness), $\kappa$, is defined as *(36)*:

$$\kappa = \sqrt{\frac{2v^2 e^2 n_b}{\epsilon k_B T}}, \qquad [S9]$$

where $e$ is the charge of the electron ($1.6 \times 10^{-19}$C), $v$ is the valency of each ionic species (here $v = 1$ for *NaCl*), $n_b$ is the electrolyte concentration in terms of the density number of ions in the bulk (0.1 *mM*), $k_B$ is the Boltzmann constant, and $T$ is the absolute temperature in degree K. The potential at the particle surface can be determined from the zeta potential values, $\xi$, as *(58)*:

$$\psi_0 = \xi\left(1 + \frac{z}{\alpha}\right)e^{\kappa z}, \qquad [S10]$$

where $z$ is the distance between the surface of the charged colloid and the slipping plane, usually taken to be about 5Å, and $\alpha$ is the Stokes radius of the particles.

The hydration repulsive potential develops due to molecular order in the adjacent and neighboring water molecules on hydrophilic surfaces. The hydration energy describes the free energy of polar adhesion between surfaces and water *(59)*, and for colloids and interfaces in an aqueous environment the hydration forces are attributed to the hydration of adsorbed counterions and ionic functional groups in the interface *(59-61)*. This superficial hydration results in a repulsive force between surfaces, which decays exponentially with a characteristic length, $\lambda \approx 1nm$ *(36,59)*. The corresponding interaction energy for spherical particles can be obtained through the Derjaguin approximation *(36,62)* as follows:

$$U_{hyd}(x) = \Delta G_{pwp} \pi \lambda \frac{D}{2} e^{\frac{-x}{\lambda}}, \qquad [S11]$$

where $\Delta G_{pwp} = 0.5 \times 10^{-3} \, J/m^2$ is the free energy of interaction between two particles in water (*i.e.*, surface energy). This short-range interaction contributes significantly to the total pair potential of silica and clay particles *(62)*.

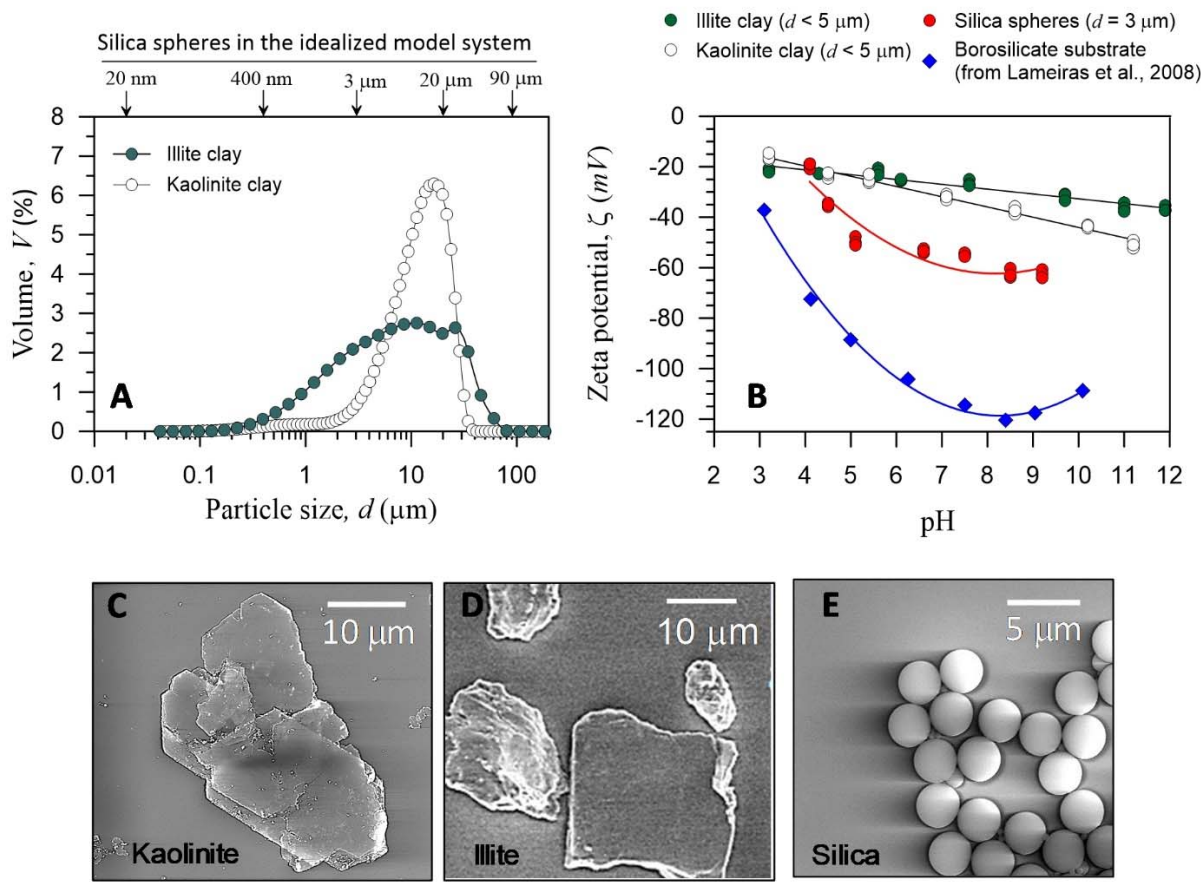

**Fig. S1. Particle size distribution and surface charge properties of the tested particles.** (**A**) Particle size distribution of unsieved illite and kaolinite clay particles measured using a laser diffraction particle sizing technique. (**B**) Bulk zeta potential of the tested particles measured at various pH conditions. The zeta potential of the borosilicate was reported from Lameiras et al. (2008) *(51)*. (**C** to **E**) SEM images showing the morphology of clay particles and silica spheres.

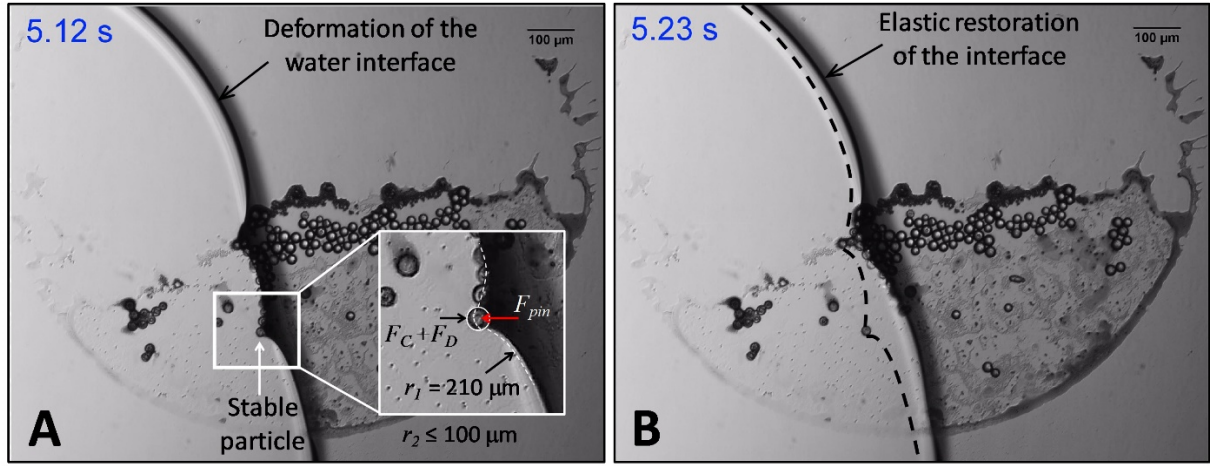

**Fig. S2. Pinning and deformation of contact line during rewetting.** (**A**) Contact line deforms, and a cusp is formed due to pinning by the particle surfaces. (**B**) The cusp breaks, and the contact line moves across the stable particle, while unstable aggregates disintegrate and their constituent particles are pushed by the contact line. $F_{pin}$ is the pinning force at the cusp location, $F_C$ and $F_D$ are the capillary and drag forces imparted on the particle. $r_1$ and $r_2$ are the radii of the water meniscus shaping the cusp around the pinning area (see Supplementary Information, *section I. Forces acting on particles during rewetting*).

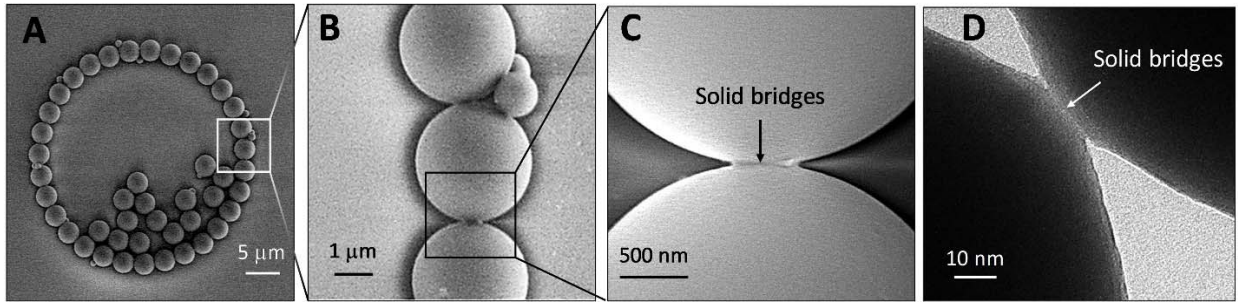

**Fig. S3. Solid bridges due to the presence of nanoparticulate contamination.** (**A** to **C**) SEM images showing the formation of solid bridges between two adjacent 3-*μm* silica particles due to the presence of submicron silica nanoparticles (identified using EDXS). (**D**) TEM image demonstrating the formation of solid bridges between two adjacent silica particles with diameter of 0.4 *μm*. The SAED data showed that the solid bridges at this scale are also composed of amorphous silica nanoparticles.

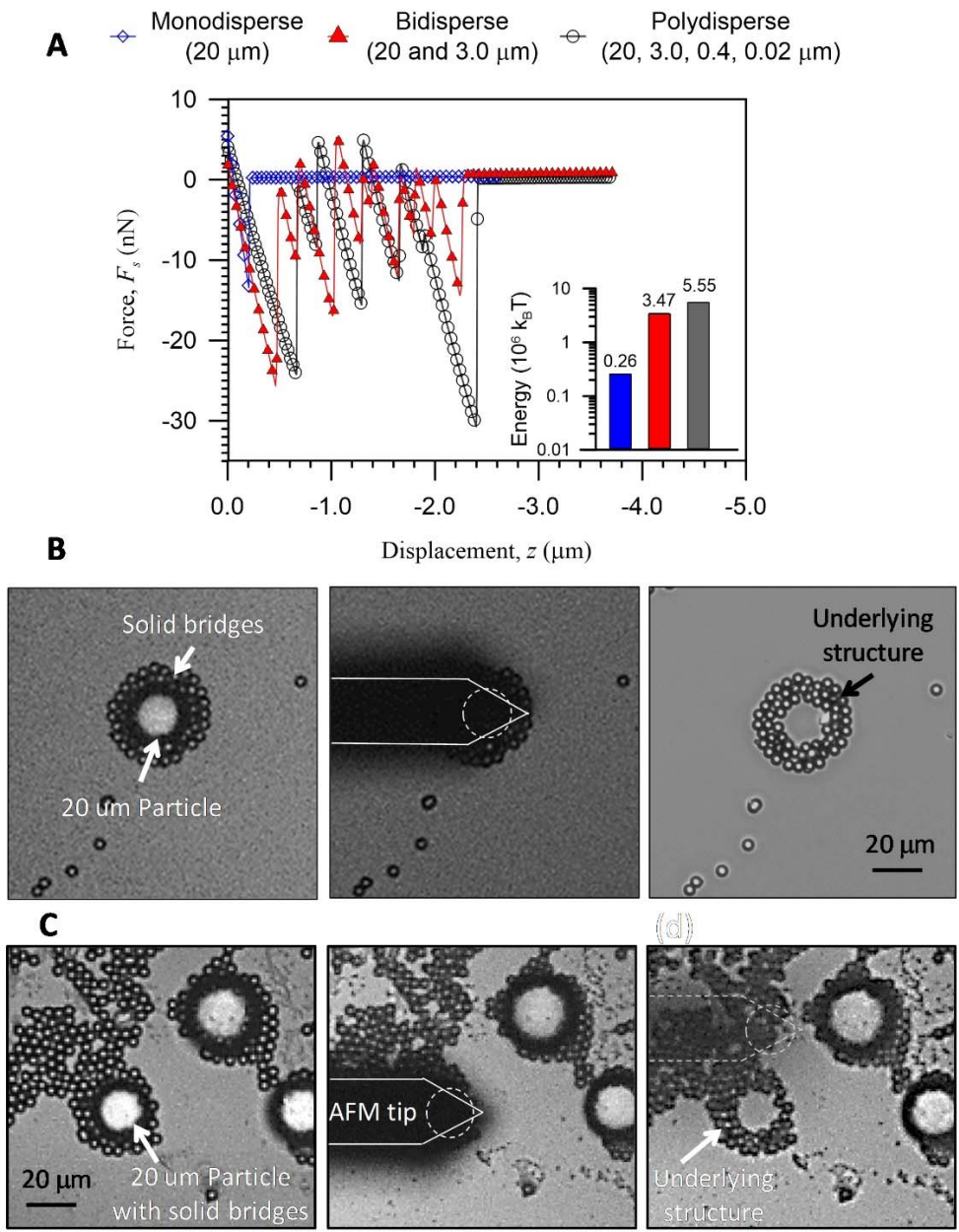

**Fig. S4. Detachment force measured using AFM technique and the underlying solid bridging structures.** (**A**) Force-displacement ($F_s$-$z$) and detachment energy (inset) of a single 20 $\mu m$ silica particle without intentional solid bridges (monodisperse suspension) and with solid bridges after evaporation of a bidisperse and polydisperse suspension. (**B** and **C**) Optical images showing the procedure for detaching a 20 $\mu m$ particle from the substrate, and the underlying solid bridges in a bidisperse and polydisperse system, respectively.

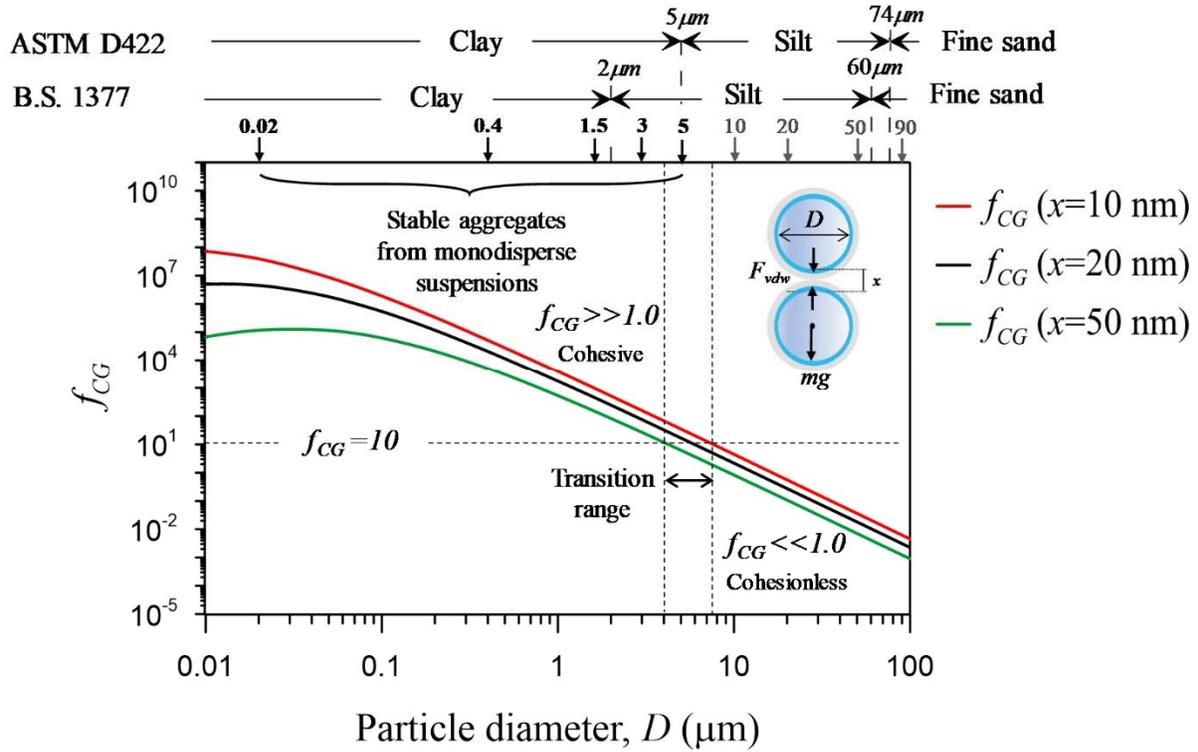

**Fig. S5. Bond-weight ratio as a function of particle size in an ideal particle to particle interaction.** $f_{CG} = \frac{U_{vdw}}{U_G}$ is the ratio of the van der Waals energy, $U_{vdw}$ (see Supplementary Information, *section II. The interfacial potential energy functions*) between two approaching particles forming aggregates and the gravitational energy, $U_G = mgD$, acting on particles, where $m$ is the particle mass, $g$ is gravitational acceleration, and $D$ is the particle diameter. $U_{vdw}$ is plotted for three typical values of surface-to-surface distance, $x$, based on our microscopic observations. When $f_{CG} = 1$, the energies balance; when $f_{CG} \ll 1$, the particle weight is dominant over the effect of the interparticle attraction and the aggregate is termed cohesionless; and when $f_{CG} \gg 1$, the system is cohesive. We experimentally probed the stability of aggregates formed by monodisperse suspensions for the range of silica spheres shown on the plot. A critical size range is suggested for the transition from cohesionless to cohesive aggregates, where $f_{CG} \approx 10$.

**Movie S1**

The movie shows a deposit formed by evaporation from a suspension of polydisperse silica spheres composed of 20, 3, and 0.4-$\mu m$ particles, subject to rewetting and transport in a microfluidic channel.

**Movie S2**

The movie shows an evaporating suspension of bidisperse silica spheres composed of 20 and 3-$\mu m$ particles on a silicate coverslip substrate. The retreating air-water interface drags 3-$\mu m$ particles and condenses some of them within the meniscus formed between the 20-$\mu m$ particles and the substrate or between adjacent particles. After water is fully evaporated, small particles form 'solid bridges' that connect larger particles to the substrate and to each other to make aggregates.

**Movie S3**

The movie shows aggregates formed by evaporation from a suspension of bidisperse silica spheres composed of 20, and 3-$\mu m$ particles subject to rewetting. The 20-$\mu m$ particles that are stabilized through solid bridging are not transported by the flow.

**Movie S4**

The movie shows a deposit formed by evaporation from a monodisperse suspension of 20-$\mu m$ particles subject to rewetting, where particles are easily transported due to the absence of solid bridges.

**Movie S5**

The movie shows aggregates formed by evaporation from a suspension of unsieved, polydisperse illite clay particles subject to rewetting. Aggregates are stable due to solid bridges.

**Movie S6**

The movie shows aggregates of illite clay modified by sieving out particles < 5 $\mu m$ subject to rewetting, and transport of particles due to the absence of small particles and the associated solid bridges.

**Movie S7**

The movie shows the rewetting behavior of aggregates formed by evaporation from a suspension of unsieved, polydisperse kaolinite clay particles. Aggregates are stable due to solid bridges.

**Movie S8**

The movie shows aggregates of same kaolinite clay modified by sieving out particles < 5 $\mu m$ subject to rewetting. Most aggregates are unstable due to the absence of solid bridges.

**Movie S9**

The movie shows water condensation on surfaces of both the substrate and silica particles during rewetting in a microfluidic channel. Hydration due to humidity precedes the wetting front and could weaken inter-particle bonds due to development of interfacial (repulsive) forces.

**Movie S10**

The movie shows water condensation on surfaces of kaolinite clay particles during rewetting in a microfluidic channel.